 \documentclass[final,1p,times]{elsarticle}

\usepackage{graphicx}

\usepackage{amssymb}
\usepackage{amsmath}

\usepackage{ulem}

\usepackage{dcolumn}

\usepackage{color}

\journal{Physics Letters B}

\begin{document}

\begin{frontmatter}



\title{$\Upsilon$ Production as a Probe for Early State Dynamics in High Energy Nuclear
Collisions at RHIC}


\author[add1]{Yunpeng Liu}
\author[add1]{Baoyi Chen}
\author[add2]{Nu Xu}
\author[add1]{Pengfei Zhuang}

\address[add1]{Physics Department, Tsinghua University, Beijing
100084, China}
\address[add2]{Nuclear Science Division, Lawrence Berkeley
National Laboratory, Berkeley, California 94720, USA}

\begin{abstract}
$\Upsilon$ production in heavy ion collisions at RHIC energy is
investigated. While the transverse momentum spectra of the ground
state $\Upsilon(1s)$ are controlled by the initial state Cronin
effect, the excited $b\bar b$ states are characterized by the
competition between the cold and hot nuclear matter effects and
sensitive to the dissociation temperatures determined by the heavy
quark potential. We emphasize that it is necessary to measure the
excited heavy quark states in order to extract the early stage
information in high energy nuclear collisions at RHIC.
\end{abstract}


\end{frontmatter}


Quarkonium production in relativistic heavy ion collisions is widely
accepted as a probe of deconfinement phase transition at finite
temperature and density. The $J/\psi$ suppression\cite{Matsui} has
been observed at SPS\cite{Topilskaya,Alessandro} and
RHIC\cite{Adare} energies and considered as a signature of the
created new state of matter, the so-called quark-gluon plasma (QGP).
At RHIC energy, however, the $J/\psi$ production and suppression
mechanisms are complicated, there are primordial production and
nuclear absorption in the initial state and
regeneration\cite{BraunMunzinger,Thews,Yan,Zhao} and anomalous
suppression during the evolution of the hot medium.

$\Upsilon$ mesons, the bound states of bottom quarks, may offer a
relatively cleaner probe into the hot and dense
medium\cite{Digal,Noronha,Vogt}. At RHIC energy, there are three
important advantages in studying $\Upsilon$ production compared with
$J/\psi$: (i) $\Upsilon$ regeneration in the hot medium can be
safely neglected\cite{Grandchamp} due to the small production cross
section of bottom quarks in nucleon+nucleon collisions. The initial
creation becomes the only production mechanism and the perturbative
QCD calculations are more reliable for estimating the production.
(ii) $\Upsilon$s are so heavy, there is almost no feed-down for
them. (iii) From the $\Upsilon$ measurement in d+Au collisions where
there is no hot nuclear matter effect, the nuclear modification
factor is about unity, implying that the cold nuclear absorption is
negligible\cite{Liu}. Therefore, the heavy and tightly bound
$\Upsilon$s can only be dissociated in the QGP phase. It is
interesting to note that the recent experimental result in Au+Au
collisions at RHIC\cite{Atomssa} has already indicated a strong
medium effect on $\Upsilon$ production.

The quarkonium dissociation temperature in the hot medium can be
determined by solving the Schr\"odinger equation for $c\bar c$ or
$b\bar b$ system with potential $V$ between the two heavy
quarks\cite{Satz}. The potential depends on the dissociation process
in the medium. For a rapid dissociation where there is no heat
exchange between the heavy quarks and the medium, the potential is
just the internal energy $U$, while for a slow dissociation, there
is enough time for the heavy quarks to exchange heat with the
medium, and the free energy $F$ which can be extracted from the
lattice calculations is taken as the potential\cite{Karsch,Shuryak}.
From the thermodynamic relation $F=U-TS<U$ where $S$ is the entropy
density, the surviving probability of quarkonium states with
potential $V=U$ is larger than that with $V=F$. In the literatures,
a number of effective potentials in between $F$ and $U$ have been
used to evaluate the charmonium evolution in QCD
medium\cite{Satz,Shuryak,Wong}.

In this Letter we investigate $\Upsilon$\ mid-rapidity production at
RHIC energy ($\sqrt{s_{NN}}=200$\ GeV), by solving a classical
Boltzmann equation for the phase space distribution of $\Upsilon$
states moving in a hydrodynamic background medium. Since the
$\Upsilon$ transverse momentum distribution should be more sensitive
to the dynamic evolution of the system, compared with the global
$\Upsilon$ yield, we will calculate not only the centrality
dependence of the nuclear modification factor, but also its
transverse momentum dependence and the averaged transverse momentum
for the ground and excited $\Upsilon$ states. We will also discuss
the dependence of these observables on the heavy quark potential and
the similarity between $\Upsilon$ production at RHIC and $J/\psi$
production at SPS.

From the experimental data in p+p collisions, $51\%$ of the observed
ground state $\Upsilon(1s)$ is from the direct production, and the
decay contributions from the excited $b\bar b$ states
$\Upsilon(1p)$, $\Upsilon(2s)$, $\Upsilon(2p)$ and $\Upsilon(3s)$
are respectively $27\%, 11\%, 10\%$ and $1\%$\cite{Affolder}. The
other two states $\eta_b(1s)$ and $\eta_b(2s)$ are scaler mesons
with typical width of strong interaction and therefore not
considered here. To simplify the numerical calculation, we will not
distinguish $\Upsilon(2p)$ from $\Upsilon(1p)$ and $\Upsilon(3s)$
from $\Upsilon(2s)$ and take the contribution fractions from the
directly produced $\Upsilon(1s), \Upsilon(1p)$ and $\Upsilon(2s)$ to
the observed $\Upsilon(1s)$ in the final state to be
$a_\Upsilon=51\%,\ 37\%$ and $12\%$.

Like the description for $J/\psi$\cite{Liuy}, the $\Upsilon$ motion
in a hot medium is characterized by the classical transport equation
\begin{equation}
\label{transport}
p^\mu\partial_\mu f_\Upsilon=-C f_\Upsilon,
\end{equation}
where $f_\Upsilon (\vec p, \vec x,t)$ is the $\Upsilon$ distribution
function in the phase space with $\Upsilon = 1s, 1p, 2s$, and the
loss term $C$ is responsible to $\Upsilon$ suppression in the hot
medium. We have neglected here the $\Upsilon$ regeneration at RHIC
energy, as discussed above. Taking into account the gluon
dissociation process $\Upsilon+g\rightarrow b+\bar b$ as the
suppression source, the loss term can be written as
\begin{equation}
\label{loss} C=\int \frac{d^3\vec k}{(2\pi)^3 E_g}F(k,p)f_g(k,T,u)\
\sigma(k,p,T),
\end{equation}
where $\vec k$ is the gluon momentum, $E_g=|\vec k|$ the gluon
energy, $F(k,p)=k_\mu p^\mu$ the flux factor, $f_g(k,T,u_\mu)$ the
gluon thermal distribution as a function of the local temperature
$T$ and velocity $u_\mu$ of the medium, and $\sigma(k,p,T)$ the
dissociation cross section at finite temperature. The cross section
in vacuum can be calculated with the Operator Production Expansion
method\cite{Peskin,Bhanot,Arleo,Oh} and is often used for $J/\psi$
suppression and should be better for $\Upsilon$ suppression. The
medium effect on the cross section is reflected in the temperature
dependence of the $\Upsilon$ binding energy $\epsilon_\Upsilon$. By
solving the Schr\"odinger equation for the $b\bar b$ system with the
heavy quark potential $V$ at finite temperature, one obtains
$\epsilon_\Upsilon(T)$ and the wave function $\psi(\vec x,T)$ and in
turn the average size of the system $\langle r\rangle(T)$. With
increasing $T$, $\epsilon_\Upsilon$ decreases and vanishes at the
dissociation temperature $T_\Upsilon$ and $\langle r\rangle$
increases and goes to infinity at $T_\Upsilon$. From the lattice
simulation\cite{Asakawa} on the $J/\psi$ spectral function at finite
temperature, the shape of the spectral function changes only a
little for $T<T_{J/\psi}$ but suddenly collapses around the
dissociation temperature $T_{J/\psi}$. To simplify our numerical
calculation, we replace the temperature dependence of the binding
energy in the cross section by a step function,
$\sigma(k,p,T)=\sigma(k,p)/\Theta(T_\Upsilon-T)$. Under this
approximation, while the cross section becomes temperature
independent at $T<T_\Upsilon$, the dissociation rate
$\alpha=C/E_\Upsilon$ depends still on the hot medium, since the
gluon density is sensitive to the temperature.

\begin{figure}[!hbt]
\centering
\includegraphics[width=1\textwidth]{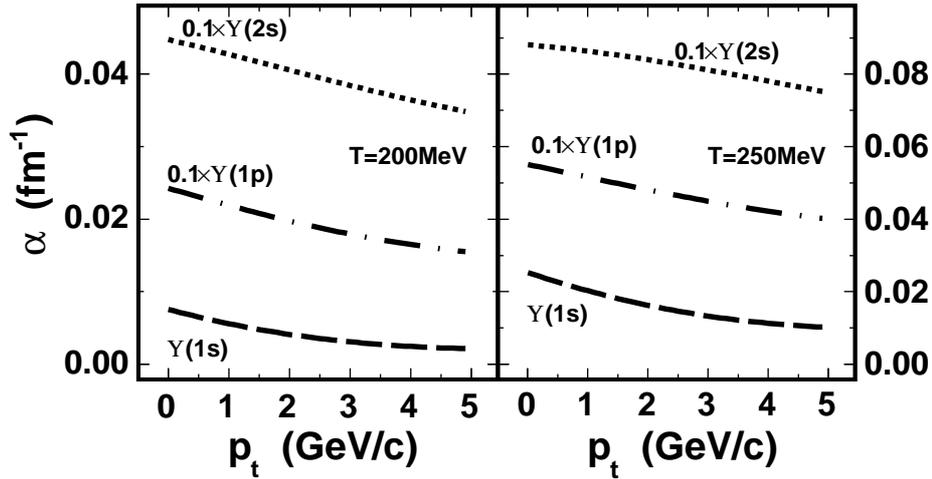}
\caption{The $\Upsilon$ dissociation rate $\alpha$ as a function of
transverse momentum at temperature $T=200$ MeV(left panel) and $250$
MeV(right panel) for the potential $V=U$. The medium velocity is
fixed as $v_{QGP}=0.5$ and its direction is chosen as the same as
the $\Upsilon$ momentum. $\Upsilon(1s), \Upsilon(1p)$ and
$\Upsilon(2s)$ are respectively shown by dashed, dot-dashed and
dotted lines. The rates for $\Upsilon(1p)$ and $\Upsilon(2s)$ are
multiplied by a factor $0.1$.}\label{fig1}
\end{figure}

For $V=U$, H.Satz and his collaborators\cite{Satz,Digal2} solved the
Schr\"odinger equation and found the dissociation temperatures
$T_\Upsilon/T_c=4,\ 1.8,\ 1.6$ for $\Upsilon=1s,\ 1p,\ 2s$,
respectively, where $T_c=165$ MeV\cite{Kolb} is the critical
temperature for the deconfinement. Since the lattice calculated
potential is mainly in the temperature region $T/T_c<4$,
$T_\Upsilon/T_c=4$ is considered as a low limit of the dissociation
temperature for the ground state. In the same way, we numerically
solved the Schr\"odinger equation for $V=F$ and the dissociation
temperatures $T_\Upsilon/T_c=3,\ 1.1,\ 1$ for $\Upsilon=1s,\ 1p,\
2s$, respectively. For all $\Upsilon$ states, the dissociation
temperatures are about $30\%$ lower in case of $V=F$ compared to
that of $V=U$. In Fig.\ref{fig1}, we show the $\Upsilon$
dissociation rate $\alpha$ as a function of transverse momentum at
fixed temperature $T_c<T<T_\Upsilon$ in the case of $V=U$. In the
calculation here we have chosen a typical medium velocity
$v_{QGP}=0.5$ and assumed that $\vec v_{QGP}$ and $\Upsilon$
momentum $\vec p$ have the same direction. All the rates are large
at low momentum and drop off at high momentum. When the temperature
increases from $200$ MeV (left panel) to $250$ MeV (right panel),
the rates for all $\Upsilon$ states increase by a factor of about
$2$. This can be understood qualitatively by the temperature
dependence of the gluon density, $n_g(T=250$ MeV$)/n_g(T=200$
MeV$)=(250/200)^3=1.95$.

In our following numerical calculations, the local temperature
$T(\vec x,t)$ and medium velocity $u_\mu(\vec x,t)$ which appear in
the gluon distribution function $f_g(k,T,u)$ and step function
$\Theta(T_\Upsilon-T)$ are controlled by the ideal hydrodynamic
equations
\begin{equation}
\label{hydro}
\partial_\mu T^{\mu\nu} = 0,\ \ \ \ \partial_\mu \left(n_B u^\mu\right) = 0,
\end{equation}
where $T^{\mu\nu}$ is the energy-momentum tensor, and $n_B$ the
baryon density. Taking into account the Hubble-like expansion
assumption for the longitudinal motion\cite{Bjorken}, the above
hydrodynamics describes the transverse evolution of the medium in
the central rapidity region. To close the hydrodynamic equations, we
take the equation of state\cite{Kolb,Sollfrank} for ideal parton gas
with parton masses $m_u=m_d=m_g=0$ and $m_s=150$ MeV and hadron gas
with hadron masses up to $2$ GeV. Here we did not consider the
back-coupling of the $\Upsilon$ states to the medium evolution. The
initial condition for the hydrodynamic equations is determined by
the corresponding nucleon+nucleon collisions and colliding nuclear
geometry\cite{Zhu}, leading to an initial temperature $T_i=340$ MeV.

The transport equation (\ref{transport}) can be solved
analytically\cite{Zhu}. The initial transverse momentum distribution
$f_\Upsilon$ can be described by the Monte Carlo event generator
PYTHIA with MSEL=62\cite{pythia} for the nucleon+nucleon collisions
and the Glauber model for the nuclear geometry. Since the collision
time for the two nuclei to pass through each other at RHIC energy is
less than the starting time of the hot medium, all the cold nuclear
matter effects can be included in the initial distribution.
Considering the fact that there is almost no $\Upsilon$ suppression
in d+Au collisions\cite{Liu}, we neglect the nuclear absorption and
take into account only the Cronin effect\cite{Gavin,Hufner}, namely
the gluon multi-scattering with nucleons before the two gluons fuse
into an $\Upsilon$. The Cronin effect leads to a transverse momentum
broadening, the averaged transverse momentum square $\langle
p_t^2\rangle_{NN}$ in nucleon+nucleon collisions is extended to
$\langle p_t^2 \rangle_{NN} + a_{gN} l$, where $l$ is the path
length that the two gluons travel in the cold nuclear medium and
$a_{gN}$ is a constant determined by the p+A data.

\begin{figure}[!hbt]
\centering
\includegraphics[width=0.8\textwidth]{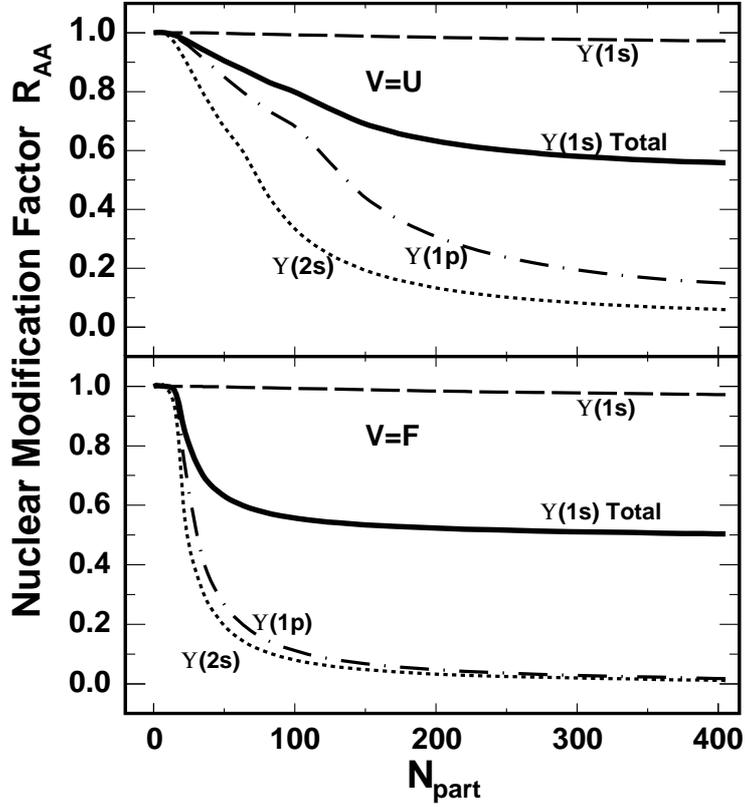}
\caption{Centrality dependence of the nuclear modification factors
$R_{AA}$ in Au+Au collisions at top RHIC energy $\sqrt {s_{NN}}=200$
GeV. Upper and lower panels are for the heavy quark potential $V=U$
and $V=F$, respectively. The directly produced $\Upsilon(1s),
\Upsilon(1p), \Upsilon(2s)$ and the total $\Upsilon(1s)$ are
respectively shown by dashed, dot-dashed, dotted and solid lines. }
\label{fig2}
\end{figure}

With the known distribution function $f_\Upsilon(\vec p, \vec x,
t)$, we can calculate the yield $N^\Upsilon_{AA}$ in the final state
by integrating $f_\Upsilon$ over the hypersurface in the phase space
determined by the critical temperature $T_c$. The nuclear
modification factors defined by
$R_{AA}=N^\Upsilon_{AA}/\left(N_{coll} N_{NN}^\Upsilon\right)$ are
shown in Fig.\ref{fig2} as functions of the number of participant
nucleons $N_{part}$ in Au+Au collisions at $\sqrt s=200$ A GeV,
where $N_{coll}$ is the number of binary collisions and
$N_{NN}^\Upsilon$ is the $\Upsilon$ yield in the corresponding
nucleon+nucleon collisions. Since the heavy quark potential at
finite temperature is not yet clear, we take the two limits of $V=U$
and $V=F$ to determine the $\Upsilon$ dissociation temperature
$T_\Upsilon$. For the ground state $\Upsilon(1s)$, the binding
energy is about 1.1 GeV which is much larger than the typical
temperature of the fireball $T\sim 300$ MeV at RHIC energy.
Therefore, $\Upsilon(1s)$ is unlikely to be destroyed in heavy ion
collision at RHIC. Considering the fact that the dissociation
temperatures $T_\Upsilon=4T_c$ for $V=U$ and $3T_c$ for $V=F$ are
both far above the temperature of the fireball, the suppression due
to the gluon dissociation is very small and independent of the heavy
quark potential. As to the excited states, the situation is
different as their binding energies are much smaller and thus their
yields are strongly suppressed. Especially, at $V=F$ which leads to
the lowest dissociation temperatures around the phase transition for
$\Upsilon(1p)$ and $\Upsilon(2s)$, almost all of them are eaten up
by the fireball in semi-central and central collisions. Including
the decay contribution from the excited states to the ground state,
the finally observed nuclear modification factor for $\Upsilon(1s)$,
$R_{AA} = \sum_{\Upsilon=1s,1p,2s}a_\Upsilon
N^\Upsilon_{AA}/\left(N_{coll} N_{NN}^\Upsilon\right)$ is controlled
by both the ground and excited states. In central Au+Au collisions,
as one can see in the figure, $R_{AA}\sim a_{1s}$=0.51. Integrating
over the number of participant nucleons, the $R_{AA}$ in minimum
bias events is 0.53 for $V=F$ and 0.63 for $V=U$. Both values are in
reasonable agreement with the preliminary PHENIX data
$R_{AA}<0.64$\cite{Atomssa}. As one can see in the figure, while the
overall $R_{AA}$ shows limited sensitivity to the heavy quark
potential, the excited states display different dependence on the
potential used in the calculation.

\begin{figure}[!hbt]
\centering
\includegraphics[width=0.8\textwidth]{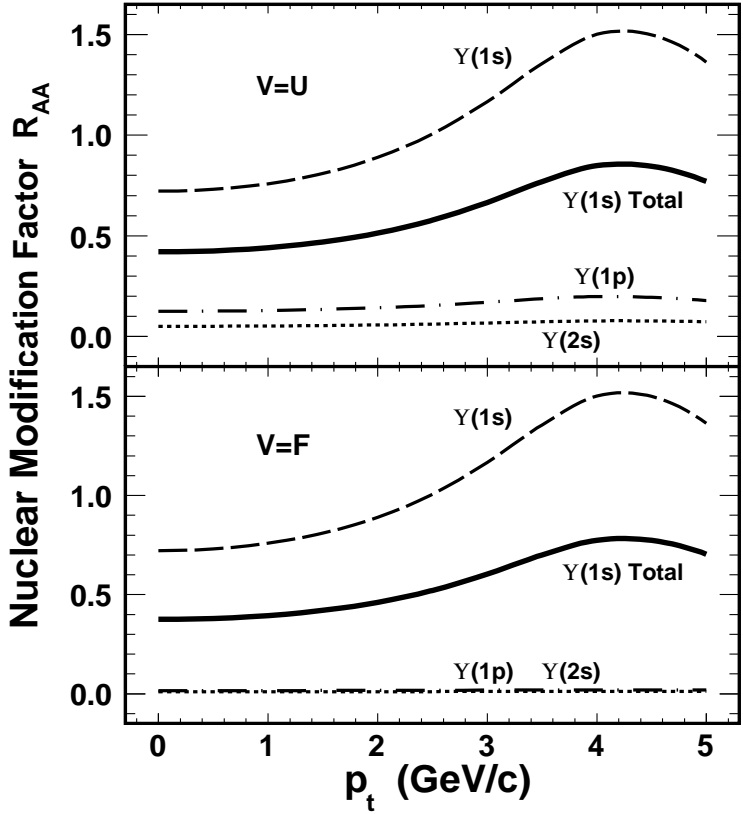} %
\caption{Transverse momentum dependence of the nuclear modification
factors $R_{AA}$ in central Au+Au collisions at top RHIC energy
$\sqrt {s_{NN}}=200$ GeV. The peak caused by the Cronin effect is
clear at $p_t\sim$ 4.5 GeV/c for both potentials.} \label{fig3}
\end{figure}

In our numerical calculations, the momentum parameters $\langle
p_t^2\rangle_{NN}$ and $a_{gN}$ are respectively taken to be 7.1
(GeV/c)$^2$ from the PYTHIA simulation\cite{pythia} and 0.2
(GeV/c)$^2$/fm to best describe the available data for p+A
collisions\cite{Wang}. In order to further address the $\Upsilon$
production dynamics, we calculated the nuclear modification factor
$R_{AA}$ as a function of transverse momentum $p_t$ in central Au+Au
collisions at RHIC energy, see Fig.\ref{fig3}. There are three
nuclear matter effects that may affect the $\Upsilon$ $p_t$
dependence: the Cronin effect in the initial state, the leakage
effect\cite{Matsui,Zhu} for higher $p_t$ particles, and the
suppression mechanism in the hot medium. While both the Cronin
effect and leakage effect lead to a $p_t$ broadening, i.e. reduction
of low $p_t$ particles and enhancement of high $p_t$ particles, a
strong enough suppression may weaken the broadening. The reason is
the following: When the gluon traveling length $l$ which
characterizes the Cronin effect is long, the temperature which
controls the suppression becomes high, the competition between the
Cronin effect and the suppression may reduce the $p_t$ broadening.
For the ground state, there is only a weak suppression, a very
strong $p_t$ broadening due to the Cronin effect and leakage effect
is expected. Since most of the excited states are dissociated in
central collisions at both $V=U$ and $V=F$, their $p_t$ broadening
is strongly suppressed and not sensitive to the heavy quark
potential. The shape of the finally observed total $R_{AA}$ is
determined by the directly produced $\Upsilon(1s)$, as shown in
Fig.\ref{fig3}.

\begin{figure}[!hbt]
\centering
\includegraphics[width=0.8\textwidth]{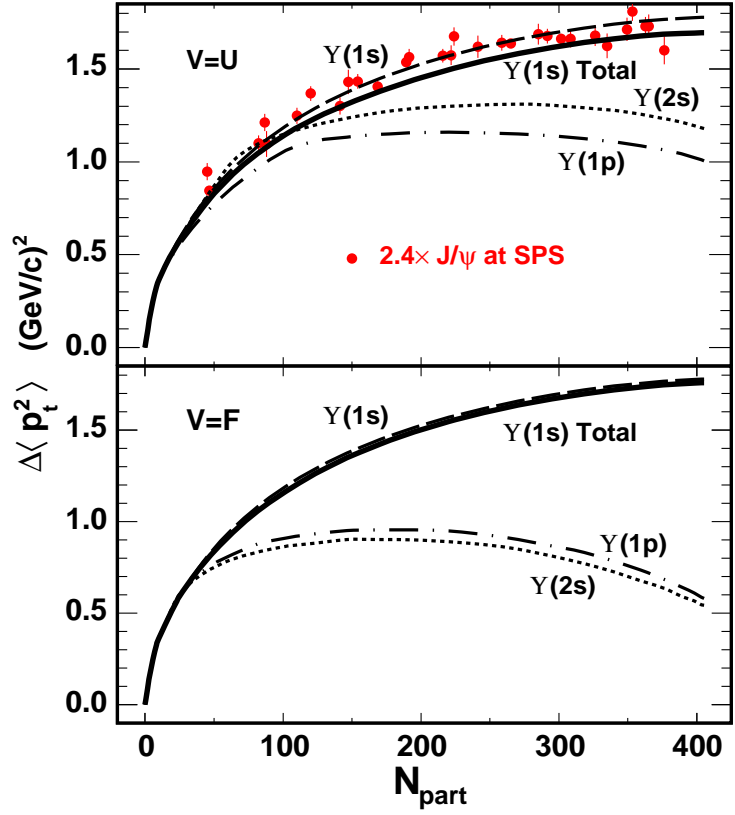}%
\caption{Centrality dependence of the difference $\Delta\langle
p_t^2\rangle\equiv\langle p_t^2\rangle_{AA}-\langle
p_t^2\rangle_{NN}$ between the averaged transverse momentum square
from Au+Au and p+p collisions at top RHIC energy $\sqrt
{s_{NN}}=200$ GeV. For comparison, we showed in the upper panel also
the $J/\psi$ data at SPS energy\cite{Topilskaya} multiplied by a
factor $2.4$.} \label{fig4}
\end{figure}

To obtain more dynamic information on the nuclear medium through
survived excited states, we calculated the averaged transverse
momentum square $\langle p_t^2\rangle_{AA}$ as a function of
centrality for Au+Au collisions, the results are shown in
Fig.\ref{fig4}. To reduce the theoretical uncertainty in $\langle
p_t^2\rangle_{NN}$ and focus on the nuclear matter effect, we
considered the difference between nucleus+nucleus and
nucleon+nucleon collisions, $\Delta\langle p_t^2\rangle\equiv\langle
p_t^2\rangle_{AA}-\langle p_t^2\rangle_{NN}$. For the directly
produced ground state $\Upsilon(1s)$, the suppression is weak and
the Cronin effect plays the dominant role. As a result,
$\Delta\langle p_t^2\rangle$ increases monotonously with collision
centrality. For the excited states, however, the effect of initial
Cronin effect is overwhelmed by the disassociation especially in
central collisions. In other words, in the most central collisions,
the high temperature region is larger than that in peripheral
collisions, most of the excited $\Upsilon$s are destroyed.
Therefore, as one can see, the value of $\Delta\langle p_t^2\rangle$
goes up in peripheral collisions due to the Cronin effect, then
becomes saturated in semi-central collisions from the competition
between the Cronin effect and the increased suppression, and finally
starts to decrease when the suppression becomes strong. In addition,
as one can see in the figure, in case of $V=F$, the decrease is
remarkable due to the stronger suppression effect. Different from
the total $R_{AA}(N_{part})$ and $R_{AA}(p_t)$ which are
approximately one half of the corresponding values for the directly
produced $\Upsilon(1s)$, see Figs.\ref{fig2} and \ref{fig3}, the
total $\Delta\langle p_t^2\rangle$ is very close to the one for the
ground state, as shown in Fig.\ref{fig4}.

It is interesting to compare the $\Upsilon$ production at RHIC
energy and the $J/\psi$ production at SPS energy. Assuming there is
no $\Upsilon$ regeneration at RHIC and no $J/\psi$ regeneration at
SPS, the production mechanism in the two cases is then characterized
by the initial creation. In addition, from the relationship between
the fireball temperature and the dissociation temperature for the
ground state at $V=U$, $T_\Upsilon=4T_c\gg T_\text{RHIC}$ and
$T_{J/\psi}=2T_c \gg T_\text{SPS}$, the suppression of
$\Upsilon(1s)$ at RHIC and $J/\psi$ at SPS is negligible. Therefore,
the averaged transverse momentum squares for $\Upsilon$ in Au+Au
collisions at RHIC and $J/\psi$ in Pb+Pb collisions at SPS are
related to each other through the Cronin effect,
\begin{equation}
\label{dpt2} \Delta\langle p_t^2\rangle|_\Upsilon^\text{RHIC} =
\frac{a_{gN}^\text{RHIC} R_\text{Au}}{a_{gN}^\text{SPS}R_\text{Pb}}\Delta\langle
p_t^2\rangle|_{J/\psi}^\text{SPS}=2.4\Delta\langle
p_t^2\rangle|_{J/\psi}^\text{SPS},
\end{equation}
where we have taken $a_{gN}^\text{SPS}=0.08$
GeV$^2$/fm\cite{Gavin,Hufner}, and $R_\text{Au}$ and $R_\text{Pb}$
are the nuclear radii. This relation in fact predicts that the
centrality dependence of $\Delta \langle p_t^2\rangle$ for
$\Upsilon$ at RHIC is proportional to that for $J/\psi$ at SPS. In
the upper panel of Fig.\ref{fig4} we showed the $J/\psi$ data at
SPS\cite{Topilskaya} multiplied by the factor $2.4$. It is clear
that the relation (\ref{dpt2}) works well. For $R_{AA}(N_{part})$
and $R_{AA}(p_t)$, their behavior depends on the details of the hot
medium, it becomes difficult to obtain similar relations between
$\Upsilon$ at RHIC and $J/\psi$ at SPS.
\begin{figure}[!hbt]
\centering
\includegraphics[width=0.8\textwidth]{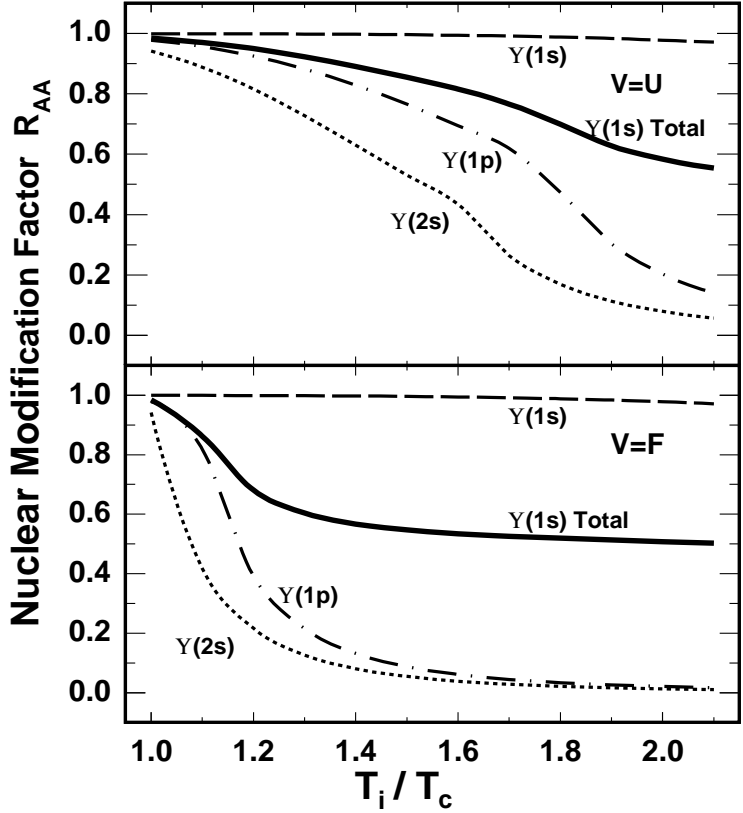}%
\caption{Initial temperature dependence of the nuclear modification
factor $R_{AA}$ in central Au+Au collisions at top RHIC energy
$\sqrt {s_{NN}}=200$ GeV. } \label{fig5}
\end{figure}

How hot is the fireball formed in relativistic heavy ion collisions?
This is a crucial question that will have an influence on all
signatures for QGP formation. In all above calculations, the initial
temperature $T_i=340$ MeV is determined by the initial energy
density and baryon density which are controlled by the
nucleon+nucleon collisions and the nuclear geometry. To extract the
initial temperature of the system from $\Upsilon$ production, we now
take $T_i$ as a free parameter and calculate the momentum integrated
$R_{AA}$ as a function of $T_i$ in central Au+Au collisions at RHIC
energy, the result is shown in Fig.\ref{fig5}. While in case of
$V=F$, see bottom plot of Fig.\ref{fig5}, the values of $R_{AA}$ are
almost constants in the temperature region $1.5<T_i/T_c<2$ which is
the expected initial temperature region for collisions at RHIC, for
$V=U$, see top plot of Fig.\ref{fig5}, the excited states and the
finally observed ground state are sensitive to the temperature.
Therefore, the experimental results of $R_{AA}$ for any state will
allow us to extract the information on the initial temperature of
the system.

In summary, we studied $\Upsilon$ production in high energy nuclear
collisions at RHIC in a transport model. The observed $\Upsilon(1s)$
is mainly from the direct production, and the contribution from the
feed down of the excited states is small. The transverse momentum
distribution of $\Upsilon(1s)$ is not sensitive to the hot medium,
but characterized by the Cronin effect in the initial stage. The
above conclusion is almost independent of the heavy quark potential.
However, the behavior of the excited $\Upsilon$ states is controlled
by the competition between the cold and hot nuclear matter effects
and sensitive to the heavy quark potential. Therefore, the yield and
transverse momentum distribution for the excited states should be
measured in the future experiments, in order to probe the dynamic
properties of the formed fireball. The initial state Cronin effect
can be studied via the centrality dependence of $\langle
p_t^2\rangle$ in A+A collisions. We did not address the influence of
the parton distribution in A+A collisions\cite{Kharzeev,Ferreiro}.
Although it will affect the details of the predictions made in this
letter, the qualitative trends, especially the nature of the high
sensitivity of the excited $\Upsilon$ states to the potential and
initial temperature, will remain to be true. The consideration on
the velocity dependence of the dissociation temperature and the
influence of reduced binding energy in low temperature region will
be discussed in the future.

\appendix
{\bf Acknowledgement:} The work is supported by the NSFC grant Nos.
10735040, 10847001, 10975084 and 11079024, and the U.S. Department
of Energy under Contract No. DE-AC03-76SF00098.











\end{document}